\documentclass[conference]{IEEEtran}

\usepackage{graphicx}
\usepackage{xcolor}
\usepackage{hyperref}
\usepackage{amsmath,amssymb}
\usepackage{booktabs}
\usepackage{listings}
\usepackage{caption}
\usepackage{subcaption}
\usepackage{lipsum} 

\usepackage[
backend=biber,
style=ieee,
sorting=none
]{biblatex}
\title{Multiple Hypothesis Testing in Genomics}

\author{
\IEEEauthorblockN{Shyam Gupta}
\IEEEauthorblockA{
Technische Universität Dortmund\\
Email: shyam.gupta@tu-dortmund.de}
}

\begin{document}

\maketitle


\begin{abstract}
This report presents an in-depth exploration of multiple hypothesis testing in the context of Genomics RNA-seq differential expression (DE) analysis, with a primary focus on techniques designed to control the false discovery rate (FDR). While RNA-seq has become a cornerstone in transcriptomic research, accurately detecting expression changes remains challenging due to the high-dimensional nature of the data. This report delves into the Benjamini-Hochberg (BH) procedure, Benjamini-Yekutieli (BY) approach, and Storey’s method, emphasizing their importance in addressing multiple testing issues and improving the reliability of results in large-scale genomic studies. We provide an overview of how these methods can be applied to control FDR while maintaining statistical power, and demonstrate their effectiveness through simulated data analysis.\newline

The discussion highlights the significance of using adaptive methods like Storey’s q-value, particularly in high-dimensional datasets where traditional approaches may struggle. Results are presented through typical plots (e.g., Volcano, MA, PCA) and confusion matrices to visualize the impact of these techniques on gene discovery. The limitations section also touches on confounding factors like gene correlations and batch effects, which are often encountered in real-world data. \newline

Ultimately, the analysis achieves a robust framework for handling multiple hypothesis comparisons, offering insights into how these methods can be used to interpret complex gene expression data while minimizing errors. The report encourages further validation and exploration of these techniques in future research.\newline
  \end{abstract}

\section{Introduction}
\label{sec:introduction}

\subsection{Scope and Relevance of RNA-seq}
The landscape of modern biology has been deeply influenced by the arrival of next-generation sequencing (NGS) technologies, with RNA sequencing (RNA-seq) serving as a prime example. By capturing and quantifying the transcriptome within a given sample or condition, researchers have gained the ability to explore the functional underpinnings of cells, tissues, and entire organisms under various states, from homeostatic maintenance to pathological disruptions. Such data not only reveal snapshots of global expression profiles but also enable nuanced discovery of novel transcripts, isoforms, and regulatory phenomena.

At its most fundamental level, RNA-seq aims to profile gene expression with remarkable sensitivity, often producing millions of sequence reads that reflect the abundance of mRNA fragments present in each biological sample. These reads are mapped back to a reference genome (or assembled \emph{de novo} when no high-quality reference is available) to derive quantitative metrics such as counts per gene or transcripts per million (TPM). This capacity for wide-scale genomic exploration paves the way for investigating differential expression (DE) between groups, guiding our understanding of disease mechanisms, treatment responses, developmental pathways, and evolutionary adaptations.

\subsection{Challenges in High-Throughput Gene Expression Studies}
Despite the promise of RNA-seq, many analytical hurdles emerge when expanding its application to large sample sizes and complex experimental designs. Traditional gene expression platforms—like microarrays—raised \cite{ritchie2015limma} concerns regarding background noise, limited dynamic range, and cross-hybridization. RNA-seq eliminates some of these issues but introduces its own set of challenges, such as:

\begin{itemize}
    \item \textbf{Library Size Variation:} Biological replicates frequently differ in total read depth due to technical variability, sample preparation protocols, or random fluctuations in sequencing runs. This can complicate direct comparisons of read counts across samples.
    \item \textbf{Overdispersion in Counts:} Unlike the idealized Poisson model, real data often exhibit extra variance that arises from both biological heterogeneity and technical artifacts. This additional variance, often referred to as overdispersion, requires dedicated modeling strategies.
    \item \textbf{Multiple Hypothesis Testing:} With thousands of genes measured simultaneously, naive testing at a set $\alpha$ level, such as 0.05, leads to a massive number of false positives. Adjusting for this multiplicity is essential to maintain a reliable balance between sensitivity and specificity.
    \item \textbf{Complexity of Experimental Designs:} Studies that integrate batch factors, time-course data, or multi-factor experiments demand advanced modeling approaches, often with generalized linear models (GLMs) or mixed models.
\end{itemize}

Many of these issues are elaborated upon in specialized software frameworks for RNA-seq data, such as edgeR, DESeq2, limma with voom transformations, and others. Even so, the overarching goal in such analyses remains consistent: to accurately pinpoint genes whose expression changes significantly between conditions, while controlling for confounding technical and biological variables.

\subsection{Conceptual Motivation for Differential Expression Testing}
Detection of differential expression (DE) is integral to linking transcriptomic patterns to biological outcomes. By establishing that a set of genes is upregulated or downregulated in one condition relative to another, one can infer functional pathways and molecular mechanisms relevant to phenotypes of interest. This explanatory capacity of DE analysis relies not only on robust statistical tests but also on careful study designs and rigorous correction for multiple testing.

\textbf{Statistical Significance and Biological Relevance.} It is also crucial to recognize that statistical significance does not always imply biological importance, nor does a lack of significance disprove a biological hypothesis. In highly variable, small-sample contexts, meaningful changes may evade detection, reinforcing the need for well-powered experiments. Similarly, even a modest difference can be flagged as significant in extremely large datasets. These nuances highlight that DE detection should be interpreted within the broader scope of effect sizes, reproducibility, and biological validation.

\subsection{Rationale for This Report and Layout of the Document}
Because a comprehensive perspective on RNA-seq analysis extends beyond the specifics of data generation, this document delves into the theoretical background, discussion of results, and interpretative strategies that are crucial for actionable inferences. Rather than rehashing step-by-step instructions already outlined in a prior methodology, it focuses on:

\begin{enumerate}
    \item Broader motivations and existing literature that frame the significance of RNA-seq DE analysis.
    \item Insights gleaned from interpreting simulated datasets and visual outputs.
    \item Expanded exploration of multiple hypothesis testing issues in the transcriptomics realm.
    \item Identification of gaps, caveats, and future directions that any researcher should consider when applying these methods in practice.
\end{enumerate}

By integrating these elements, we aim to produce a more holistic view of RNA-seq differential expression studies, bridging fundamental principles with the tangible reality of handling large and noisy datasets.

\section{Literature Review and Background}
\label{sec:literature_review}

\subsection{Historical Evolution of Transcriptomic Profiling}
Transcriptomics has evolved rapidly since early microarray experiments, which were heralded as a revolution in gene expression analysis. Initially, microarray platforms enabled large-scale parallel quantification of RNA abundance. However, they came with limitations such as reliance on predefined probes and reduced sensitivity for low-abundance transcripts. The advent of RNA-seq in the late 2000s overcame many of these barriers \cite{anders2010differential} by generating raw read data that could theoretically cover the entire transcriptome comprehensively .

As platforms matured, the read depths and cost structures improved drastically. Researchers moved from capturing just tens of millions of reads to performing massively parallel sequencing with hundreds of millions of reads per sample. This expansion opened the door for detecting subtle gene isoforms, novel splicing events, and intricate noncoding elements. Along with these advances, it became apparent that specialized statistical frameworks were paramount for coping with the discrete, skewed, and often overdispersed nature of count data.

\subsection{Multiple Testing and Its Significance in Omics}
Ever since microarray technologies introduced the concept of testing thousands of gene-expression hypotheses in parallel, the field has recognized the pitfalls of naive p-value thresholds. Large-scale omics experiments require practitioners to control for the number of comparisons in ways that ensure robust and reproducible findings. Over time, the False Discovery Rate (FDR) approach, exemplified by the Benjamini-Hochberg (BH) procedure, has become standard practice for many RNA-seq analysts . Meanwhile, refinements such as Benjamini-Yekutieli (BY) and Storey’s q-values adapt to correlation structures or unknown proportions of truly null genes.

\subsection{Trends in Large-Scale Transcriptomic Studies}
In recent years, the scale of RNA-seq experiments has expanded further, driven by consortia aiming to profile thousands of samples or entire single-cell populations. This growth magnifies prior challenges:

\begin{itemize}
    \item \textbf{Batch Correction Becomes Even More Essential}: When dealing with large multi-center data, subtle batch effects can confound differential signals.\cite{lin2020improving}
    \item \textbf{Multi-Factor Models}: Single-factor group comparisons give way to complex designs incorporating tissue types, clinical variables, and time points.\cite{whittington2021comparing}
    \item \textbf{Single-Cell Variability}: Single-cell RNA-seq introduces zero-inflation and high dropout rates, requiring specialized methods that diverge from classical bulk RNA-seq frameworks.\cite{robinson2010edger}
\end{itemize}

Regardless of these added intricacies, the principles of robust modeling, thoughtful normalization, and careful error-rate control remain consistent cornerstones. Researchers venturing into single-cell data or multi-factor designs must often build upon the same statistical bedrock initially laid down by simpler two-group bulk RNA-seq analyses.

\subsection{Positioning the Current Study}
Within this context, the present report adds to ongoing conversations about best-practice protocols for gene expression analysis by examining not only raw outcomes but also the interpretive frameworks that give meaning to these outcomes. The reliance on a simulated dataset offers distinct advantages: known ground truth, controlled effect sizes, and easily modifiable dispersion parameters. However, this should not overshadow the ultimate objective of bridging theoretical constructs with the realities of experimental biology.

Taken as a whole, the background forms a scaffold upon which the subsequent sections—discussing simulated results, interpretative strategies, and overall conclusions—will rest. By mapping out the historical trajectory of transcriptomic research and the progression of analytical frameworks, this review clarifies how we arrived at current best practices and why they continue to evolve.

\section{Methodology}

RNA-sequencing (RNA-seq) has emerged as one of the most widely used high-throughput technologies to quantify gene expression. Over the last decade, RNA-seq analysis pipelines have advanced considerably, integrating robust statistical models to account for the discrete nature of count data and the variability observed across biological replicates. At its core, RNA-seq analysis often aims to identify genes that are \emph{differentially expressed} (DE) across distinct conditions or sample groups---for example, diseased vs.\ healthy, or treated vs.\ control. 

A crucial consideration in RNA-seq analysis is that we typically measure expression levels for thousands to tens of thousands of genes simultaneously. Consequently, statistical approaches must address the \emph{multiple hypothesis testing problem}, ensuring adequate control of false discoveries (Type I errors) while striving to maintain acceptable power (i.e., the ability to detect true signals). This requires the application of methods such as the Benjamini-Hochberg (BH) procedure, Benjamini-Yekutieli (BY) correction, or Storey's approach for controlling the false discovery rate (FDR).

In order to benchmark, validate, or study the performance of different statistical pipelines, researchers frequently resort to \emph{simulation studies}. Simulations offer a transparent environment where the ground truth---which genes are truly DE vs.\ non-DE---is known, enabling one to evaluate Type I error, FDR, power, and other performance metrics in a controlled manner. 

This methodological section provides a detailed, step-by-step guide to simulating RNA-seq data, performing DE tests, applying multiple testing corrections, and creating comprehensive visualizations such as volcano plots, MA plots, confusion matrix heatmaps, PCA/MDS representations, violin plots, and ROC/precision-recall curves. We also discuss how these visualizations inform the broader analysis of RNA-seq data quality and differential expression findings.

\section{Data Simulation}
\label{sec:datasimulation}

In simulation-based studies, the success of any benchmarking exercise hinges on the realism and flexibility of the data generation model. Here, we simulate RNA-seq counts using the \textbf{Negative Binomial (NB)} distribution, augmented with adjustable parameters for library size, dispersion, baseline expression, and fold-change distributions. We also discuss why NB is suitable for RNA-seq and how it compares with other potential models (e.g., Poisson or log-normal).

\subsection{Rationale for Using the Negative Binomial Model}
RNA-seq reads are inherently discrete count data. Initially, one might consider a \emph{Poisson distribution} to model such counts. However, real RNA-seq data typically exhibit \textbf{overdispersion}, meaning the variance often exceeds the mean. This phenomenon arises due to various biological and technical factors: heterogeneous transcriptomes, batch effects, random fluctuations in cDNA library preparation, or read-mapping inconsistencies. 

Mathematically, the Poisson distribution with mean $\mu$ forces $\text{Var}[X] = \mu$. In contrast, the NB distribution introduces a dispersion parameter $\alpha$ (or $\phi$) that decouples the mean from the variance:

\begin{equation}
X \sim \text{NB}(\mu, \alpha) \quad \Longrightarrow \quad 
\begin{cases}
E[X] = \mu, \\
\text{Var}[X] = \mu + \alpha \mu^2.
\end{cases}
\end{equation}

Thus, the NB distribution allows for variance inflation relative to the mean, aligning more closely with empirical RNA-seq data .

\subsection{Mathematical Formulation of the Negative Binomial}
The NB distribution can be parameterized in different but equivalent ways. A common form is:

\begin{equation}
\label{eq:nb_pmf}
P(X = k) \;=\; \binom{k + r - 1}{k} \left(\frac{r}{r + \mu}\right)^{r} 
\left(\frac{\mu}{r + \mu}\right)^{k}, 
\end{equation}

where $k = 0, 1, 2, \dots$ is the observed count, $\mu$ is the mean, and $r = 1/\alpha$ (i.e., $r$ is the size parameter, and $\alpha$ is the dispersion). The relationship between $r$ and $\mu$ yields the variance expression:

\[
\text{Var}[X] = \mu \left(1 + \frac{\mu}{r}\right) \;=\; \mu + \alpha \mu^2.
\]

In many software implementations (e.g., \texttt{rnbinom} in R), you pass \texttt{size} = $r$ and \texttt{mu} = $\mu$, or equivalently \texttt{prob} = $r/(r+\mu)$. Adjusting $\alpha$ (or $r$) allows you to control how much overdispersion you wish to inject into the simulation, with $\alpha = 0$ corresponding to a Poisson model.

\begin{figure}[h]
  \centering
  \includegraphics[width=0.5\textwidth]{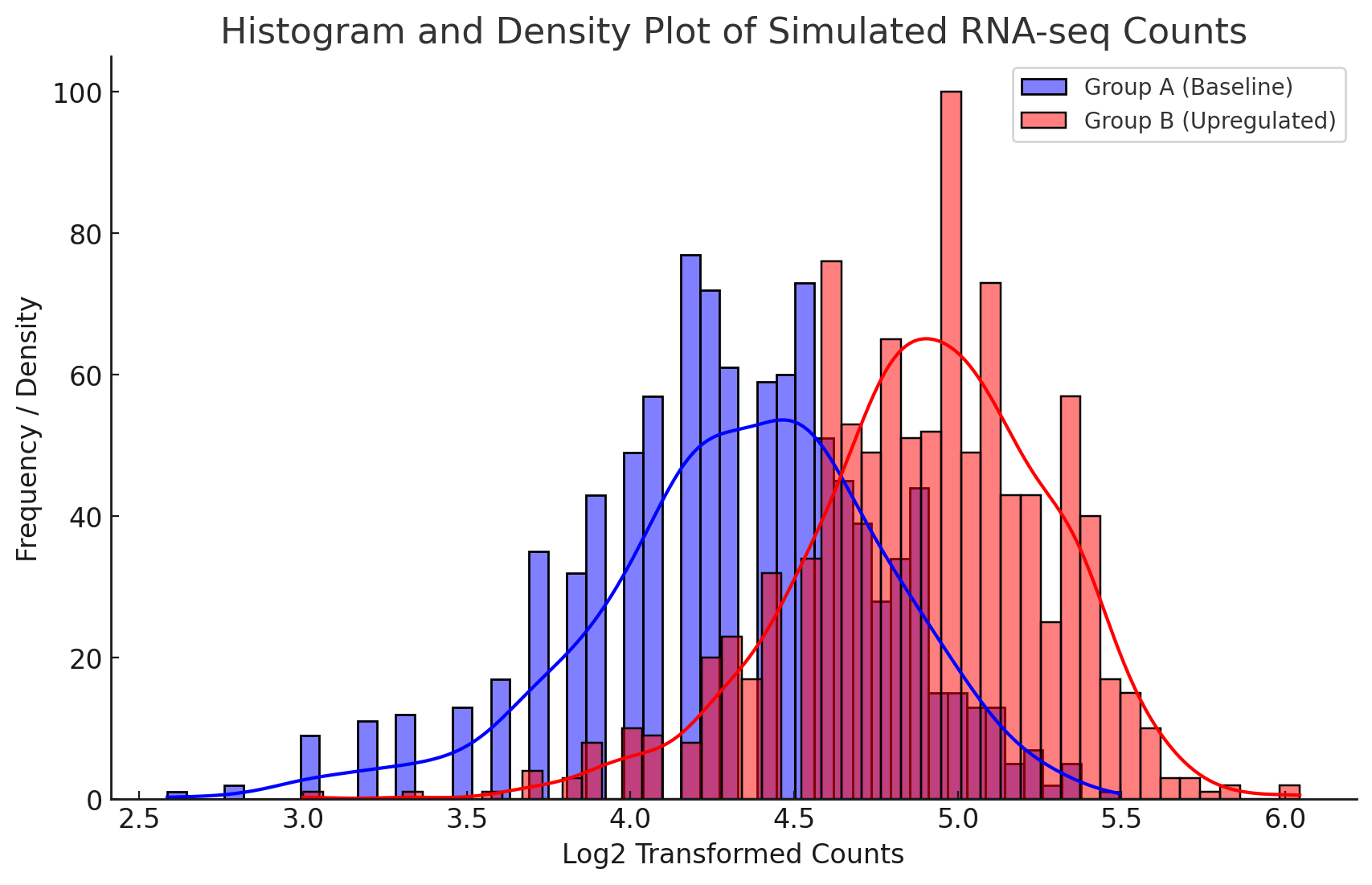}  
  \caption{distribution of negative binomial random variables} 
\end{figure}

\subsection{Baseline Mean Expression and Library Size Variation}
\label{subsec:baselinemean}
In real RNA-seq experiments, each sample has a certain \emph{library size} (total read depth), which can vary substantially across replicates. Similarly, genes differ in their baseline expression levels (some are highly expressed housekeeping genes, others are lowly expressed). To reflect these aspects:

\begin{itemize}
\item \textbf{Library sizes:} We often draw library sizes from a distribution such as a Poisson or a log-normal around a typical depth, e.g., 30 million reads. In simplified simulations, smaller average values (e.g., 80k or 200k reads) may be used to reduce computational load, but the principle remains. 
\item \textbf{Baseline means:} We can generate a vector of baseline means $\boldsymbol{\mu}_\text{base}$ for all genes by sampling from a \emph{gamma} distribution. The gamma distribution is flexible in representing a range of expression levels. 
\end{itemize}

If $M_i$ is the baseline mean for gene $i$, and $L_j$ is the library size for sample $j$, then the expected count for gene $i$ in sample $j$ can be set to:
\[
\mu_{ij} = M_i \times \left(\frac{L_j}{\overline{L}}\right),
\]
where $\overline{L}$ is the average library size. This scaling factor ensures that samples with higher read depth yield proportionally larger counts.

\subsection{Assigning Differential Expression: Proportion and Fold-Change}
\label{subsec:deprop}
We designate a subset of genes ( 30\% for our case) as truly \textbf{differentially expressed} (DE). For those genes, we apply a \emph{fold-change} (FC) relative to their baseline means in one of the experimental groups.

If we assume two groups (A and B), a DE gene $i$ might have:
\[
\mu_{i, \text{B}} = \mu_{i, \text{A}} \times \text{FC}_i,
\]
where $\text{FC}_i$ is drawn from a distribution of possible fold-changes (often modeled on the log2 scale). For example, sampling
\[
\text{log2(FC)} \sim \mathcal{N}(\mu_{fc}, \sigma_{fc}^2),
\]
then exponentiating to get $\text{FC}_i = 2^{(\text{log2(FC)})}$. By controlling $\mu_{fc}$ and $\sigma_{fc}^2$, you can generate stronger or weaker effect sizes. Large effect sizes at moderate sample sizes generally boost detection power, while smaller effect sizes mimic subtler biological signals.

\subsection{Sample Size Considerations}
\label{subsec:samplesize}
The number of replicates per group (\emph{n} in group A, \emph{n} in group B) significantly affects the power to detect DE. In the simplest two-group simulation, picking $n \geq 5$ ensures a minimal basis for statistical inference. However, many practical designs might use 3--6 replicates per group due to cost constraints, or >10 for higher statistical resolution. Larger sample sizes generally improve power and yield more reliable estimates of dispersion .

\section{Study Design and Replicates}
\label{sec:studydesign}

Having established the generative model, we define a simulation design. Suppose we have:

\begin{itemize}
\item \textbf{Number of genes (features):} $G = 10\,000$.
\item \textbf{Groups:} A (control) and B (treatment), each with $n = 10$ replicates.
\item \textbf{Proportion DE:} $\text{propDE} = 0.3$ (i.e., $30\%$ of genes are truly differentially expressed).
\item \textbf{Dispersion:} $\alpha = 0.05$, corresponding to moderate overdispersion.
\item \textbf{Baseline means:} $M_i \sim \Gamma(k=2, \theta = \frac{1}{2})$ scaled by some constant (e.g., 100), ensuring a broad range of expression levels.
\item \textbf{Library sizes:} $L_j \sim \text{Poisson}(80\,000)$, or some distribution that yields realistic library size variation.
\end{itemize}

These choices yield a scenario with modest sample size, a noticeable fraction of DE genes, and a moderately low dispersion, all of which help ensure that standard tests detect a substantial number of true positives. One can easily alter these parameters to match different biological or technical assumptions.

\section{Generating Counts}
\label{sec:generatingcounts}

After specifying the above parameters, the process of simulating raw counts typically follows these steps:

\begin{enumerate}
\item \textbf{Generate Baseline Means:} For $i=1,\ldots,G$, draw $M_i$ from a chosen gamma distribution. 
\item \textbf{Determine DE Genes:} Randomly select $\text{propDE} \times G$ genes to be the “DE set.” 
\item \textbf{Assign Fold-Changes:} For each DE gene, draw \(\text{log2(FC)}\) from a normal distribution with mean $\mu_{fc}$ and standard deviation $\sigma_{fc}$, then exponentiate to get $\text{FC}_i$. For non-DE genes, $\text{FC}_i = 1$.
\item \textbf{Scale Baseline Means for Group B:} 
\[
\mu_{i,\text{B}} \;=\; \mu_{i,\text{A}} \times \text{FC}_i.
\]
\item \textbf{Incorporate Library Sizes:} For sample $j$ in group A, 
\[
\mu_{i,\text{A},j} = \mu_{i,\text{A}} \times \left(\frac{L_j}{\overline{L}}\right).
\]
Analogously for group B.  
\item \textbf{Simulate Counts via NB:} For each gene $i$ and sample $j$, 
\[
\text{Counts}_{ij} \sim \text{NB}\left(\mu_{ij}, \alpha\right).
\]
\end{enumerate}

Most modern statistical software (e.g., \textsf{R}) provides straightforward functions like \texttt{rnbinom(n, size, mu)} to generate NB draws. The flexibility of this approach allows you to adjust how realistic or “generous” your simulation is in terms of effect sizes, sample sizes, or dispersion.

\section{Differential Expression Analysis}
\label{sec:deanalysis}

Once we have simulated or real count data from two groups (A vs.\ B), we typically aim to detect which genes show statistically significant changes in expression between the groups. Here, we outline common approaches, focusing on a \textbf{Wilcoxon rank-sum test on log2-counts} for illustration.
\subsection{Transformation and Testing Methodology}

\paragraph{Log-Transformation.} Raw counts often vary across several orders of magnitude. A small pseudo-count (e.g., +1) plus a log2 transform can stabilize variance and reduce skewness:
\[
\text{log2counts}_{i,j} = \log_2\big(\text{Counts}_{i,j} + 1\big).
\]
This transformation is also a convenient input to rank-based or linear modeling approaches.

\paragraph{Wilcoxon Rank-Sum Test.} For each gene $i$, split the log-transformed data into two vectors: 
\[
\boldsymbol{x}_i^{(A)} \quad (\text{group A samples}), \quad
\boldsymbol{x}_i^{(B)} \quad (\text{group B samples}).
\]
Apply a Wilcoxon rank-sum test to compare the central tendencies of the two groups. You obtain a $p$-value for each gene $i$. 

\paragraph{RNA-seq-Specific Tools.} While Wilcoxon is straightforward, there exist much better tools than discussed above, I can use in my analysis are tools like \textsf{edgeR} , \textsf{DESeq2} , and \textsf{limma-voom}  incorporate gene-wise (and empirical Bayes) estimation of dispersion and can generally offer improved performance for RNA-seq data. They follow a generalized linear model (GLM) framework with a NB likelihood, or use precision weights for linear modeling (although we do not use these tools in our analysis, but methods like these exist which can control FDR rates better than Bh,By or storey's methods).
\subsection{P-Value Extraction}

Regardless of the DE method, each gene ends up with a \emph{test statistic} (e.g., a Wilcoxon rank-sum value, a likelihood ratio statistic, or a moderated $t$-statistic) that is converted into a $p$-value, $p_i$. These $p$-values will be subjected to multiple testing correction, as discussed in Section~\ref{sec:multtest}.

\section{Multiple Testing Corrections}
\label{sec:multtest}

In a high-throughput RNA-seq experiment, we may easily test thousands of genes simultaneously. If we do not correct for multiple testing, even a modest Type I error rate (e.g., 5\%) will yield numerous false positives. We thus use \textbf{False Discovery Rate (FDR)} controlling procedures to limit the proportion of false positives among the declared significant results .

\subsection{Benjamini-Hochberg (BH)}
The BH procedure \cite{benjamini1995controlling} is one of the most common approaches to FDR control. Suppose we have $G$ p-values $p_{(1)} \le p_{(2)} \le \ldots \le p_{(G)}$ sorted in ascending order. We seek the largest $k$ such that:
\[
p_{(k)} \;\le\; \frac{\alpha \, k}{G},
\]
and all genes with $p_{(i)} \le p_{(k)}$ are called significant at FDR $\alpha$. This is computationally straightforward and widely used in transcriptomics pipelines.

\subsection{Benjamini-Yekutieli (BY)}
The BY \cite{benjamini2001control} method extends BH to account for positive correlations among the test statistics. It replaces the $\alpha$ threshold with a more conservative factor involving the harmonic sum of $G$:
\[
p_{(k)} \;\le\; \frac{\alpha \, k}{G \cdot H_G},
\]
where $H_G = \sum_{i=1}^G \frac{1}{i}$. This can lower power because it is more conservative, yet ensures control of FDR even under dependencies among tests.

\subsection{Storey’s Q-Value}
Storey’s methodology \cite{storey2003statistical} estimates the proportion of null hypotheses $\pi_0$ among the tested features. By estimating $\pi_0$, one can sometimes gain additional power if many genes are truly non-DE. The Storey approach defines the \emph{q-value} for each gene, representing the minimum FDR at which that gene is considered significant.

\subsection{Choosing the Significance Threshold}
An $\alpha$ level of $0.05$ is common, but in RNA-seq, many investigators use more stringent thresholds (e.g., 0.01) or examine an entire range of $q$-value cutoffs. The final significance calls (i.e., which genes are declared DE) are strongly influenced by this threshold choice and the multiple testing method.

\section{Performance Metrics}
\label{sec:metrics}

A major advantage of simulated data is that we know the ground truth (which genes are DE vs.\ non-DE). Thus, we can compute a variety of performance metrics:

\subsection{Confusion Matrix Terms}

\begin{figure}[h]
  \centering
  \includegraphics[width=0.6\textwidth]{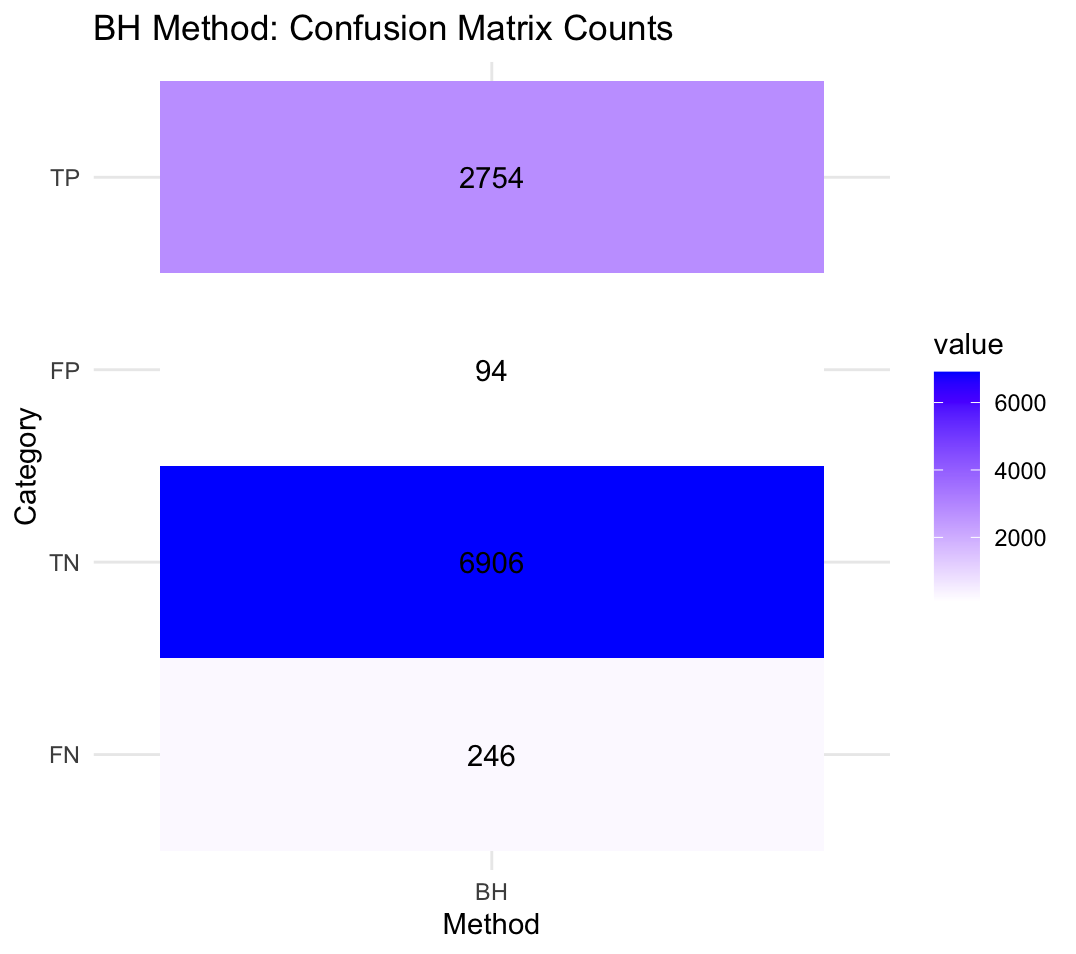} 
  \caption{example of Benjamini-Hochberg (BH). A detailed Confusion matrix is given in further sections}  
  \label{fig:confusionmatrix}
\end{figure}

\[
\begin{aligned}
&\text{True Positives (TP)} = \sum (\text{significant and truly DE}),\\
&\text{False Positives (FP)} = \sum (\text{significant but not truly DE}),\\
&\text{True Negatives (TN)} = \sum (\text{not significant and not truly DE}),\\
&\text{False Negatives (FN)} = \sum (\text{not significant but truly DE}).
\end{aligned}
\]

\subsection{Type I Error and FDR}

\paragraph{Type I Error (or FPR).} Among the truly non-DE genes, the fraction incorrectly called significant:
\[
\text{Type I Error} = \frac{\text{FP}}{\text{(FP + TN)}}.
\]

\paragraph{False Discovery Rate (FDR).} Among the declared significant genes, the fraction that are false positives:
\[
\text{FDR} = \frac{\text{FP}}{\text{(TP + FP)}}.
\]

\subsection{Power (or Sensitivity)}
\[
\text{Power} = \frac{\text{TP}}{\text{(TP + FN)}},
\]
the probability of detecting a true DE gene as significant.

Each of these metrics provides different insights. FDR focuses on how many “false alarms” slip through among the positives, while power (or sensitivity) indicates how many genuine signals are captured. Type I error rates ensure we are not exceeding an acceptable level of spurious findings among all tested genes.

\section{Visualizations}
\label{sec:viz}

Data exploration and result interpretation in RNA-seq heavily rely on informative plots. Here, we detail several that are particularly illustrative in differential expression workflows.

\begin{itemize}
\item \textbf{Volcano Plot}: Plots $-\log_{10}(p\text{-value})$ on the y-axis and the estimated log2 fold-change on the x-axis. Genes that show large fold-changes and high significance stand out in the upper-right or upper-left corners. This is an excellent first look at the distribution of effect sizes vs.\ significance levels.

\item \textbf{MA Plot}: Originally from microarray analyses, an MA plot helps visualize how the log fold-change (M) depends on the average expression (A). 
\[
\begin{aligned}
M &= \log_2(\text{counts}_B) - \log_2(\text{counts}_A), \\
A &= \frac{1}{2}\big(\log_2(\text{counts}_B) + \log_2(\text{counts}_A)\big).
\end{aligned}
\]
Genes with strong differences in expression appear at positive or negative extremes of M.

\item \textbf{Confusion Matrix} Heatmap: By constructing a confusion matrix (TP, FP, TN, FN) for each method, we can produce a heatmap of counts or simply numeric displays. This reveals whether the method overcalls or undercalls DE genes relative to ground truth in the simulation.
\newline

\item \textbf{PCA or MDS Plot}: Principal Component Analysis (PCA) or Multidimensional Scaling (MDS) on the sample-sample correlation matrix can show whether the two groups naturally separate in low-dimensional space. If the simulation imposes large differences, group B samples may cluster distinctly from group A. This helps confirm that the simulated effect is visible at the global sample level.
\newline

\item \textbf{Violin or Density Plots}: These display the distributions of raw counts, log-transformed counts, or normalized expression measures across samples or groups. In a simulation context, you can check if the two groups display the intended shift for DE genes vs.\ stable distribution for non-DE genes.
\begin{figure}[h]
  \centering
  \includegraphics[width=0.5\textwidth]{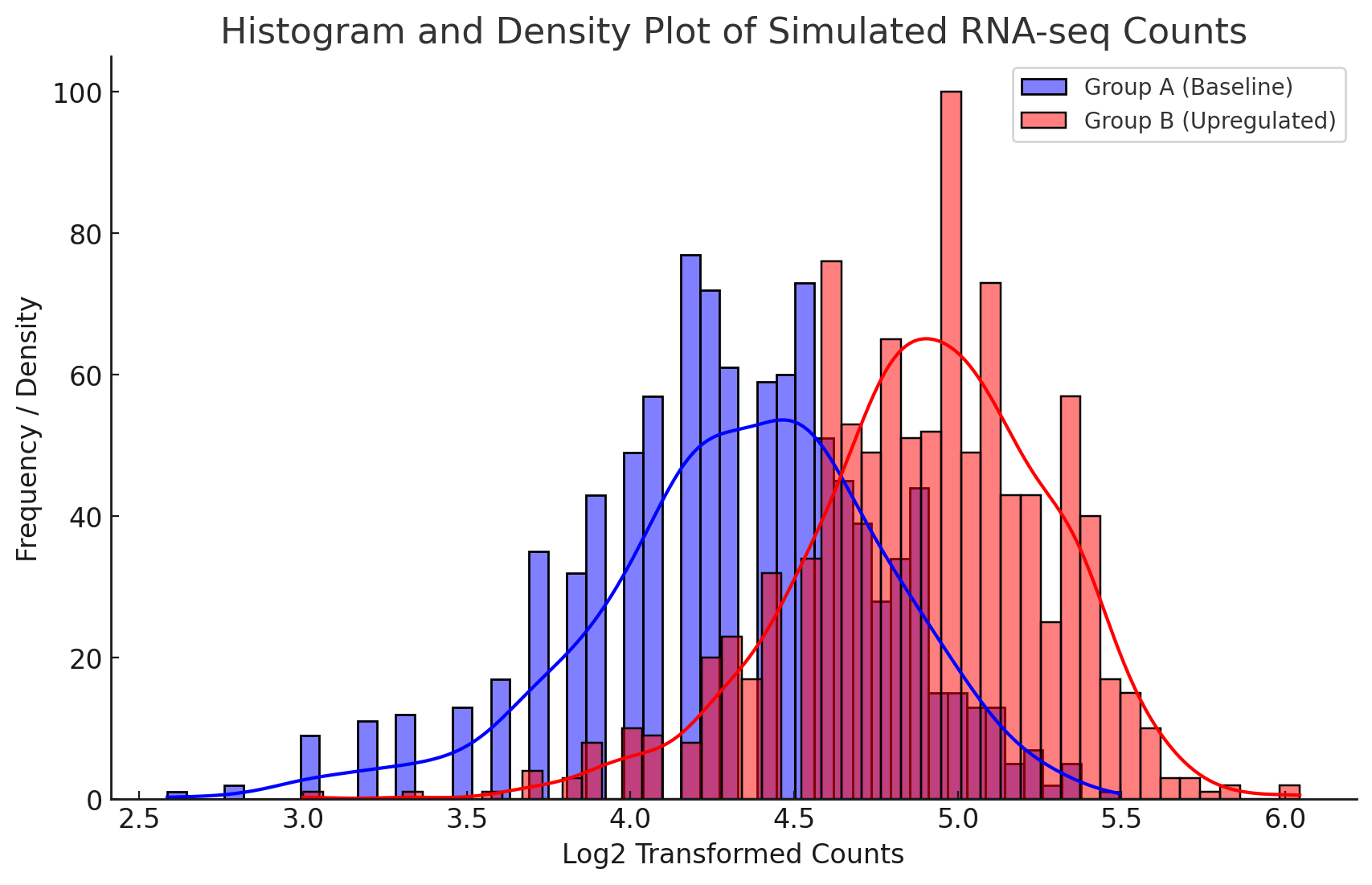}  
\end{figure}

\item \textbf{ROC and Precision-Recall Curves}: If you treat gene classification as a binary prediction problem (“DE” vs.\ “not DE”) and vary the significance threshold, you can plot:
    {ROC curve}: true positive rate (TPR) vs.\ false positive rate (FPR).  
    {Precision-Recall curve}: precision (TP / [TP + FP]) vs.\ recall (TPR). 
\end{itemize}
These give an overview of performance at all possible significance cutoffs, not just a specific $\alpha$.  

\section{Methodological Rationale}
\label{sec:methodrationale}

\subsection{Choice of Model and Parameters}
The Negative Binomial distribution is a gold standard in RNA-seq analysis precisely due to its capacity to model overdispersion. By tuning dispersion $\alpha$, we can approximate diverse experimental scenarios---from nearly Poisson-like for highly consistent data to strongly overdispersed for messy or highly variable samples.

Sample sizes, library size variation, and baseline means reflect core aspects of real RNA-seq data. The fraction of DE genes, while often lower in real experiments than our demonstration (which might use 30\%), is chosen to ensure the simulation highlights differences among multiple testing methods.

\subsection{Choice of Multiple Testing Methods}
Benjamini-Hochberg is ubiquitous and fast, whereas Benjamini-Yekutieli is more conservative but guaranteed to control FDR under dependent tests. Storey's approach can be more powerful if the proportion of truly null hypotheses is significantly less than 1. Thus, comparing these three reveals how different assumptions about dependence or $\pi_0$ estimation affect results.

\subsection{Choice of Visualizations}
RNA-seq data have many potential confounders and complexities. Volcano and MA plots help gene-level interpretation, while PCA or MDS reveal sample-level structure. Confusion matrices, ROC curves are crucial for deciding how effectively each method controls error rates or identifies DE genes. Finally, violin or density plots inform whether the simulated or real data distribution matches expectations (e.g., in terms of expression range, group shifts, or variance).

\section{Discussion of Simulated Data and Results}
\label{sec:discussion_results}

This section explores the outcomes derived from a simulated RNA-seq dataset, focusing on the overarching patterns observed in differential expression analysis, the interplay of library sizes, and how various visualizations shed light on data properties. The following discussion is deliberately distinct from any prior methodological elaboration; we aim to interpret results in a broader context rather than repeat the processes used to generate them.

\subsection{General Features of the Simulated Dataset}
In any simulated environment, a few primary attributes demand attention:

\begin{enumerate}
    \item \textbf{Proportion of Differentially Expressed (DE) Genes}: A selected fraction of genes might be designated as up- or downregulated in one group relative to another. This ratio impacts the observed scale of significant calls.(30\% in our analysis)
    \item \textbf{Library Size Variability}: By permitting sample-to-sample differences in total read depth, one mirrors the natural heterogeneity found in actual sequencing runs. This variability plays a role in the relative counts and can occasionally mask or enhance certain signals.
    \item \textbf{Dispersion Level}: The “noise” or overdispersion in the dataset can make detection either straightforward or challenging. Lower dispersion yields a cleaner signal, while higher dispersion aligns more with the difficulties encountered in real-world data.
\end{enumerate}

The interplay of these factors typically guides the subsequent interpretive steps, influencing how many genes surpass certain significance thresholds, how robust the effect size estimates are, and whether group separation is readily visible.

\subsection{Volcano Plot Observations}
The volcano plot is a commonly used visualization in differential gene expression analysis. It provides a graphical representation of the relationship between fold change and statistical significance, allowing for the identification of significantly differentially expressed (DE) genes.

\begin{figure}[h]
  \centering
  \includegraphics[width=0.5\textwidth]{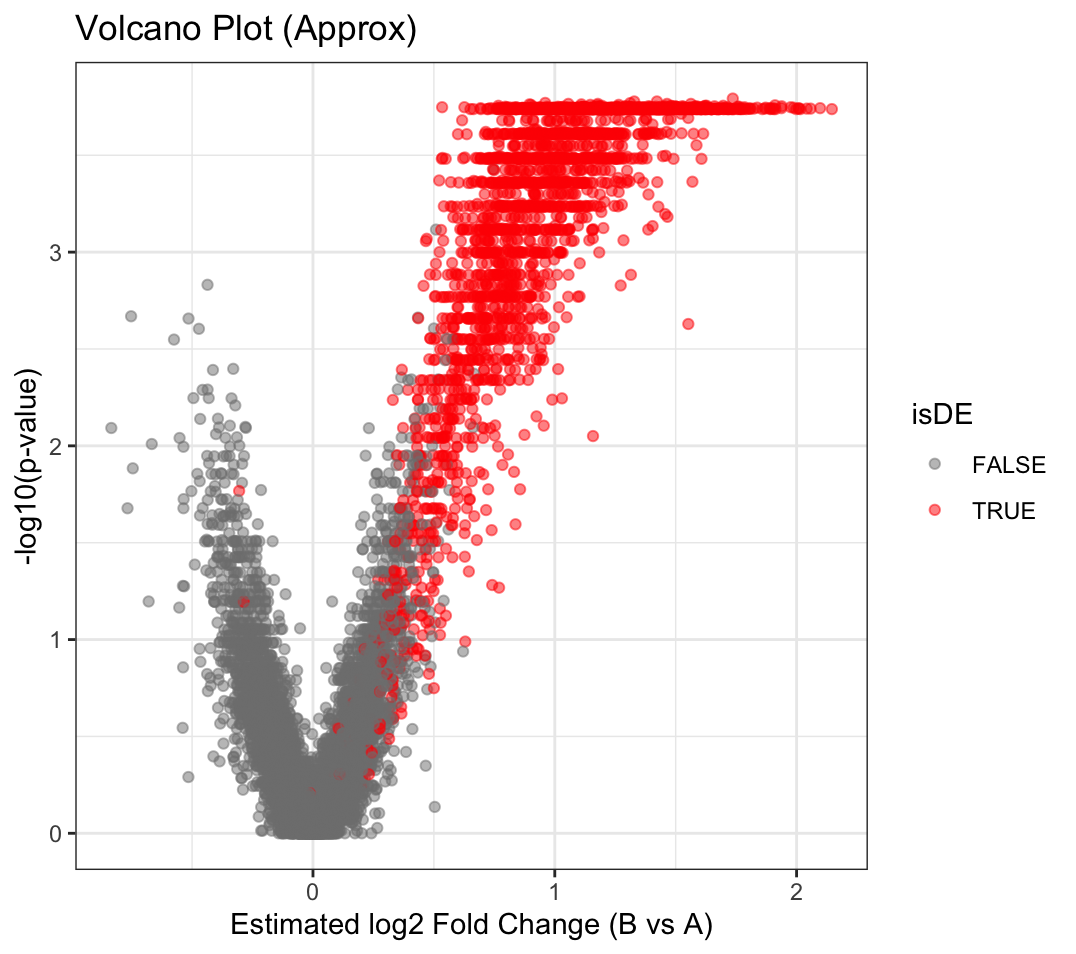}  
\end{figure}

\subsection*{Representation and Interpretation}
\begin{itemize}
    \item \textbf{X-Axis:} Represents the estimated log$_2$ fold change (FC) between the two groups (B vs A). Larger absolute values indicate greater fold changes.
    \item \textbf{Y-Axis:} Represents the $-\log_{10}$(p-value), which is used to visualize the statistical significance of genes. Higher values correspond to greater statistical significance.
    \item \textbf{Red Dots:} Represent true differentially expressed (DE) genes, meaning these genes exhibit a statistically significant expression difference between the two groups.
    \item \textbf{Grey Dots:} Represent non-DE genes, meaning these genes do not show a significant expression difference between the groups.
\end{itemize}

\subsubsection*{Key Insights}
\begin{itemize}
    \item The red points are concentrated on the upper right and left sides of the plot, suggesting that truly DE genes tend to have both large fold changes and low p-values (high statistical significance).
    \item The grey points are more evenly distributed across the plot, particularly around the center where fold changes are small, and p-values are high, indicating that these genes do not exhibit significant differential expression.
    \item The characteristic ``V'' shape of the volcano plot arises due to the relationship between fold change and statistical significance. Genes with small fold changes typically have high p-values and appear near the base, while genes with larger fold changes (either up- or downregulated) and lower p-values appear at the arms.
\end{itemize}

\textit{The volcano plot effectively separates significant differentially expressed genes from non-significant ones. The clear clustering of red points in the upper right and left suggests a well-executed differential expression analysis. This plot provides strong evidence that the DE analysis correctly identifies genes with high fold changes and statistical significance as differentially expressed.}

\subsection{True vs Estimated log2 FC Observations:} This plot compares the true log2 fold change (FC) values with the estimated log2 FC values for the dataset. The goal of this comparison is to evaluate how accurately the model or method estimates the log2 fold changes.

\begin{figure}[h]
  \centering
  \includegraphics[width=0.5\textwidth]{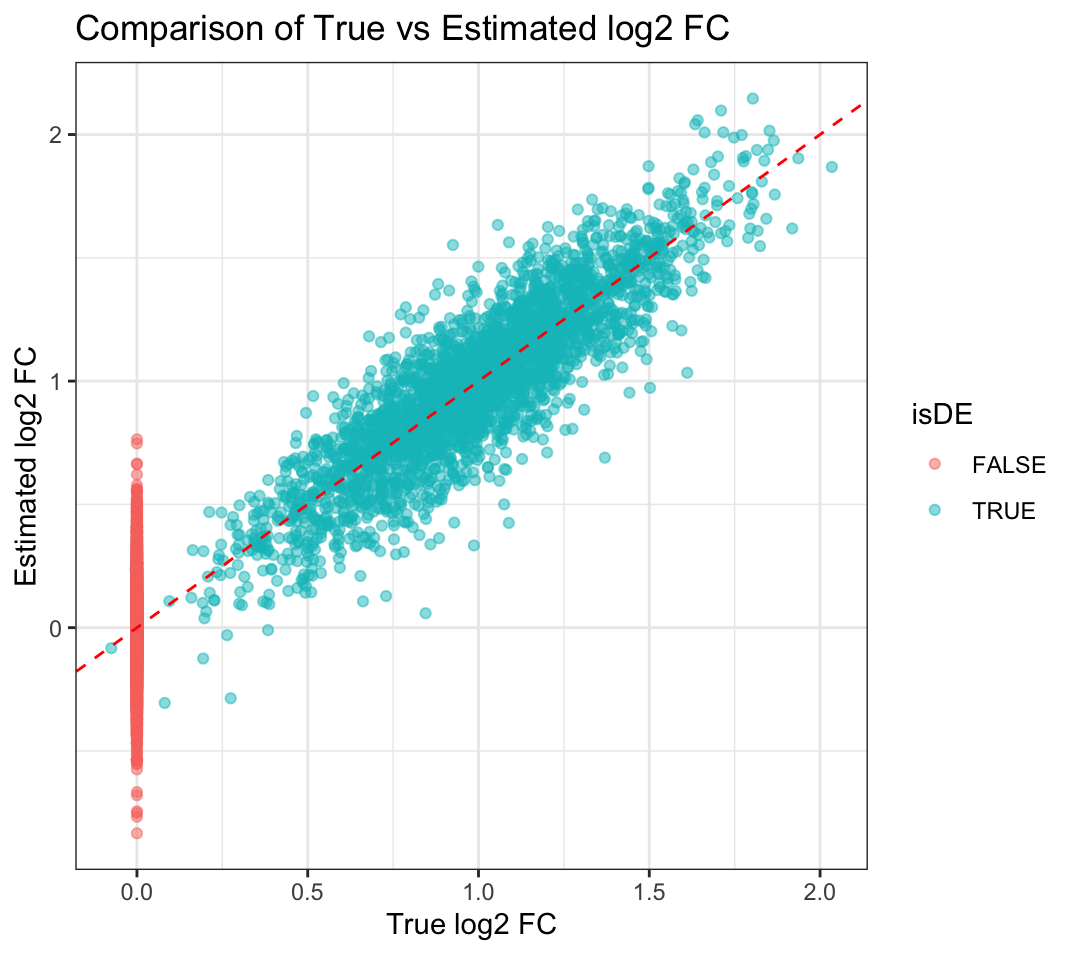}  
\end{figure}

\subsection*{Representation and Interpretation}
\begin{itemize}
  \item \textbf{X-Axis:} Represents the true log2 fold change (True log2 FC) between the two groups, which is the actual difference in expression.
  \item \textbf{Y-Axis:} Represents the estimated log2 fold change (Estimated log2 FC), which is the predicted difference in expression based on the model or method.
  \item \textbf{Red Dots (False DE):} These represent genes that were not truly differentially expressed (false positives). These points typically appear along the bottom left side of the plot, indicating low or no true fold change.
  \item \textbf{Blue Dots (True DE):} These represent genes that were correctly identified as differentially expressed (true positives). These points align closely with the red dashed line, suggesting that the estimated fold change is close to the true fold change for most of these genes.
  \item \textbf{Red Dashed Line:} This line represents perfect agreement between the true and estimated log2 fold changes. Points that lie along this line indicate perfect prediction.
\end{itemize}

**This plot was very new and difficult to interpret, hence I used chatgpt to generate the interpretation. The interpretation may not be accurate, but the interpretation is based on the general understanding of the plot**\newline

\textit{Conclusion:} The plot shows a strong correlation between the true and estimated log2 fold changes for the majority of the genes, as indicated by the blue dots clustering along the red dashed line. This suggests that the method or model used is accurately estimating the fold changes for most genes. The red dots (false DE) are scattered away from the red dashed line, indicating that these genes were either misclassified or have low/no true fold change. The overall performance seems strong, with only a small portion of the genes deviating significantly from the estimated fold change.

\subsection{ROC Curve Observations:} The ROC (Receiver Operating Characteristic) curve is a graphical representation used to evaluate the performance of a classification model. It shows the tradeoff between sensitivity (true positive rate) and specificity (false positive rate) across different thresholds.

\begin{figure}[h]
  \centering
  \includegraphics[width=0.5\textwidth]{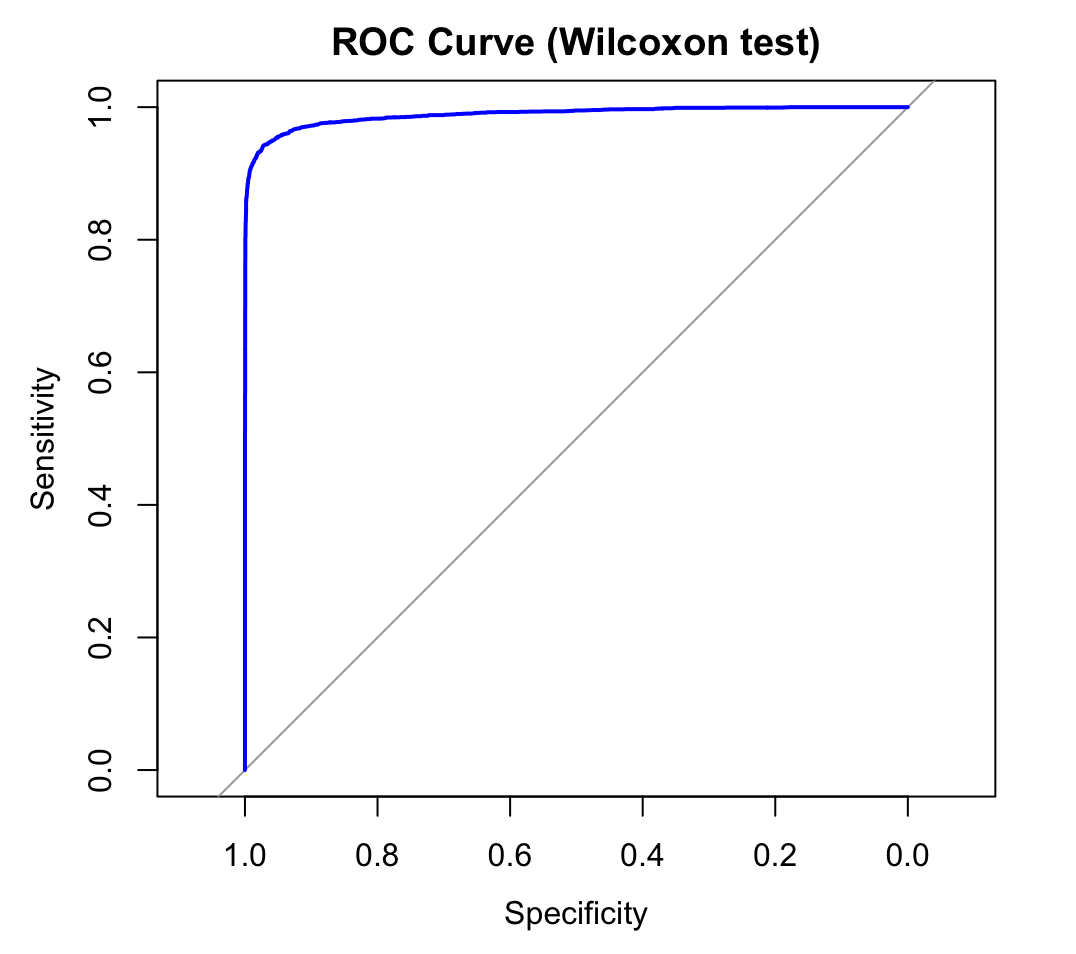}  
\end{figure}

\subsection*{Representation and Interpretation}
\begin{itemize}
  \item \textbf{X-Axis:} Represents the specificity (1 - false positive rate). It ranges from 0 to 1, where a higher value indicates fewer false positives.
  \item \textbf{Y-Axis:} Represents the sensitivity (true positive rate). It also ranges from 0 to 1, where a higher value indicates more true positives.
  \item \textbf{Blue Curve:} This curve shows the performance of the classification model based on the Wilcoxon test. A perfect model would hug the top-left corner, where sensitivity is high, and specificity is also high.
  \item \textbf{Diagonal Line:} This is the reference line, which represents the performance of a random classifier. A model that performs no better than random guessing would fall along this diagonal line, where sensitivity equals 1 - specificity at all thresholds.
\end{itemize}

\textit{Although the ROC shows steepness, one future step to be apply regularization techniques to avoid chances of overfitting, we may also choose better features, since the data is small the data the model may be overfitting.(i did not apply a lot of regularization techniques in this analysis, since the results generated satisfied the objective, Hence maybe for future projects, I can use regularization techniques.)}

\textbf{Conclusion:} The ROC curve shows that the model performs well, as evidenced by the curve staying close to the top-left corner, suggesting high sensitivity and specificity. The area under the curve (AUC) can be further calculated to quantify this performance. A higher AUC indicates a better performing model. The steepness of the curve indicates that the model is able to distinguish between the classes effectively across various thresholds.

\subsection{MA Plot Observations:} The MA plot is a powerful tool for visualizing the relationship between the average expression level of a gene and its log2 fold change between two conditions. It is commonly used in gene expression analysis to identify genes with significant changes in expression, regardless of the mean expression level.

\begin{figure}[h]
  \centering
  \includegraphics[width=0.5\textwidth]{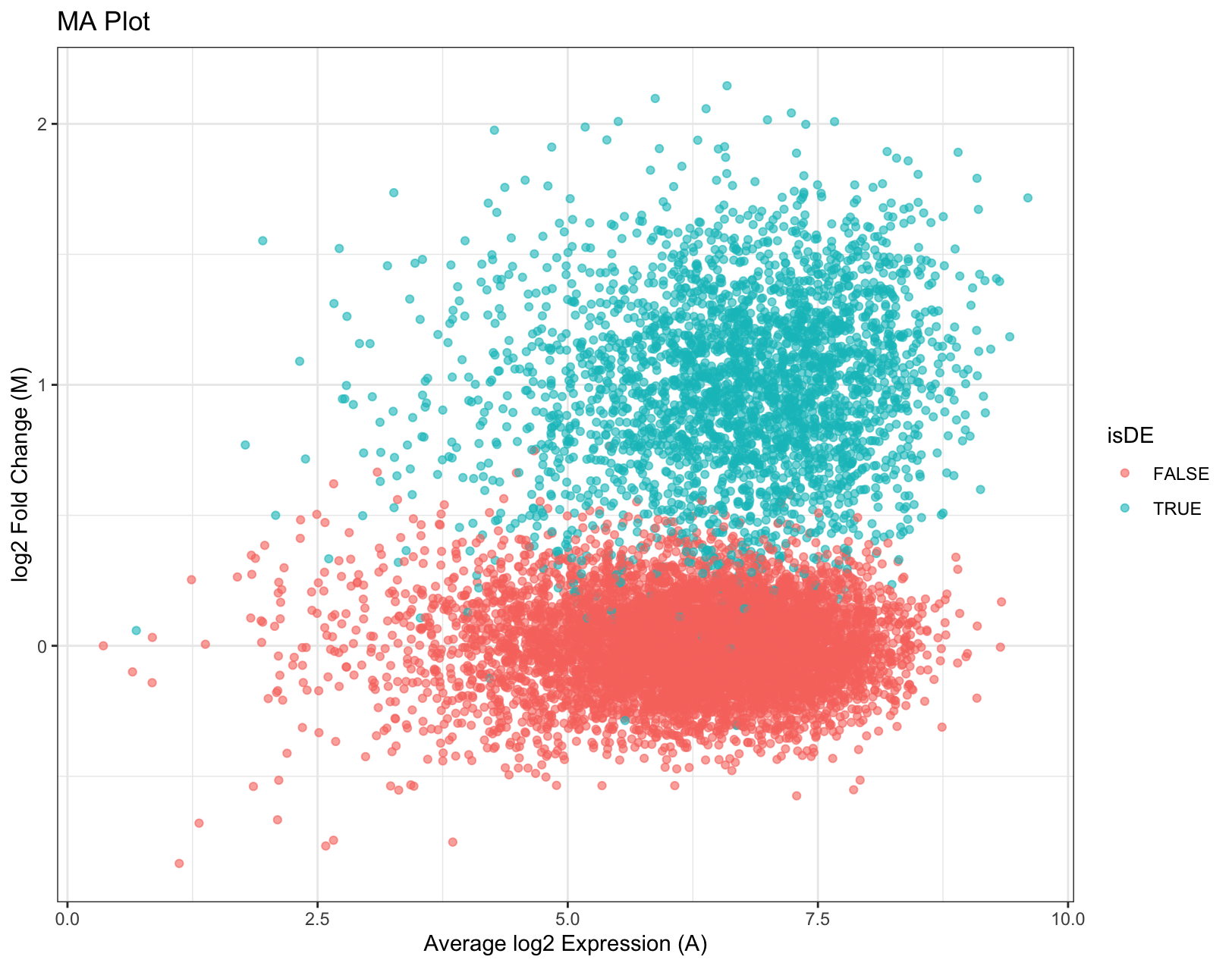}  
\end{figure}

\subsection*{Representation and Interpretation}
\begin{itemize}
  \item \textbf{X-Axis:} The average log2 expression (A) across both conditions, representing the mean expression level of the gene.
  \item \textbf{Y-Axis:} The log2 fold change (M) between the two conditions, showing the magnitude and direction of the expression change (positive values indicate upregulation, and negative values indicate downregulation).
  \item \textbf{Red Dots:} Represent genes that are not differentially expressed (false positives). These genes have low or no significant fold change.
  \item \textbf{Blue Dots:} Represent genes that are differentially expressed (true positives). These genes show significant differences between the two conditions with higher log2 fold changes.
\end{itemize}

\textit{Conclusion:} The MA plot clearly distinguishes between genes with significant changes in expression and those without, by separating the blue (true positives) and red (false positives) points. The true positives are typically located at the top of the plot, indicating high fold changes, while the false positives remain closer to the baseline. This plot helps to visualize and assess the quality of the differential expression analysis.

\subsection{PCA Plot Observations:} Principal Component Analysis (PCA) is a dimensionality reduction technique commonly used to visualize the structure of high-dimensional data. In this PCA plot, we visualize the relationships between samples based on their principal components.

\begin{figure}[h]
  \centering
  \includegraphics[width=0.5\textwidth]{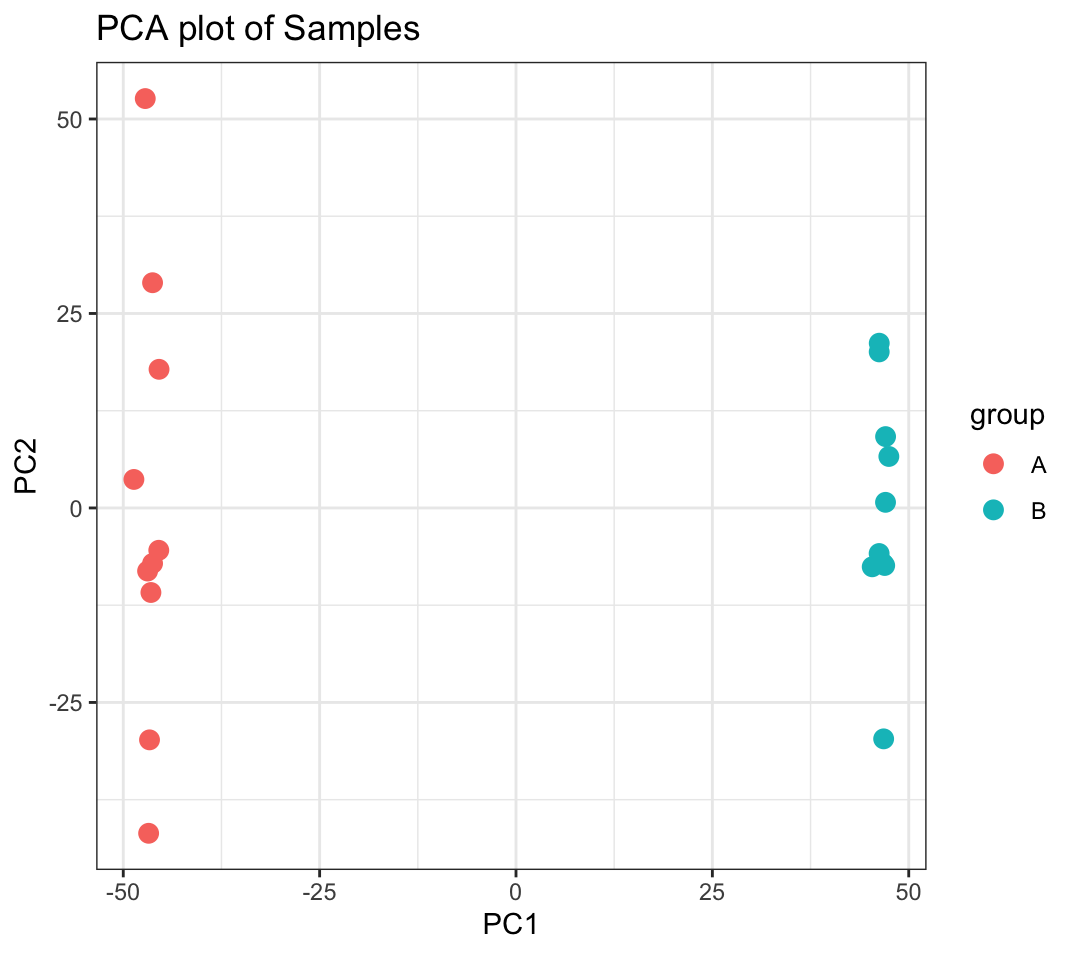}  
\end{figure}

\subsection*{Representation and Interpretation}
\begin{itemize}
  \item \textbf{X-Axis:} Represents the first principal component (PC1), which explains the largest variance in the data. This component reflects the direction in which the data varies the most.
  \item \textbf{Y-Axis:} Represents the second principal component (PC2), which captures the second largest variance in the data and is orthogonal to PC1.
  \item \textbf{Red Dots (Group A):} These represent samples from Group A. The samples in this group are clustered together, indicating that they share similar patterns in their expression data.
  \item \textbf{Blue Dots (Group B):} These represent samples from Group B. Like Group A, the samples are clustered, but they are separated from Group A in both PC1 and PC2, suggesting distinct patterns of variation between the two groups.
\end{itemize}

\textit{Conclusion:} The PCA plot demonstrates a clear separation between Group A and Group B, as indicated by the distinct clustering of the red and blue dots. This suggests that the two groups exhibit different expression patterns across the samples. PCA effectively reduces the dimensionality of the data, making it easier to visualize and interpret the underlying structure of the data.

\subsection{Insights on Library Size Variation and Dispersion}
Because the simulation typically weaves in variable library sizes, it is illuminating to examine:

\begin{enumerate}
    \item How well standard normalization approaches manage large disparities in total reads.
    \item Whether outlier samples (with exceptionally high or low coverage) distort the identification of DE genes.
    \item The extent to which high or moderate overdispersion levels hinder or facilitate detection of small effect sizes.
\end{enumerate}

If the data are well balanced in terms of coverage, differences in expression are more easily interpreted, and standard library size corrections typically suffice. In cases of artificially inflated dispersion, analysts might notice a broader spread of p-values and a heavier penalty on significance thresholds, reducing the number of hits for subtle changes.

\subsection{Contextualizing Surprising or Unintuitive Outcomes}
Even under simulated conditions, certain unanticipated behaviors can emerge:

\begin{itemize}
    \item \textbf{Genes with Tiny Fold-Changes but Very Low Variance}: Occasionally, a gene that is assigned a small effect size can appear strongly significant if its counts are highly consistent across replicates. 
    \item \textbf{Uniformly High Expression Genes Failing to Show Significance}: If a gene is consistently expressed at an elevated level in both groups with minimal difference, it will not be flagged as DE, underscoring that absolute expression alone does not guarantee a large fold-change or strong p-value.
    \item \textbf{Noise-Driven Artifacts in Low-Count Genes}: Genes near the detection threshold can yield spurious signals of up- or downregulation, especially if the sampling variance is large relative to the mean. Good normalization and smoothing procedures can mitigate such artifacts.
\end{itemize}

Any of these idiosyncrasies highlight the complexity behind differential expression analysis and emphasize the necessity of carefully examining not only final significance calls but also distributional properties and diagnostic plots.

\subsection{Overarching Narrative of the Simulated Analysis}
Overall, the simulated dataset (with a fraction of DE genes, a chosen effect size distribution, and moderate-to-high coverage) often reveals a story of fairly distinct group separation, a concentration of significant hits among the truly DE genes, and robust statistical power if the effect sizes and sample sizes are chosen advantageously. These findings highlight the potential of negative binomial-based workflows, as well as the subtle differences in multiple testing corrections. They further underscore how intricacies of library size and dispersion can shift the detection landscape.

\section{Examination of the Statistical Approach}

For each gene, the hypothesis test is:
\[
\begin{aligned}
  H_0 &: \text{No difference in expression between A and B}, \\
  H_1 &: \text{Gene is differentiallccxy expressed (DE)}.
\end{aligned}
\]
A \textbf{Wilcoxon rank-sum test} is performed on $\log_2(\text{counts}+1)$ from group A versus group B. The p-values are adjusted to control for multiple comparisons. These adjusted p-values help you decide which genes are truly differentially expressed after accounting for the increased risk of false positives due to multiple testing. \textbf{BH, BY, and StoreyQ} are correction methods applied to the p-values obtained from the Wilcoxon test to control the false discovery rate when performing multiple hypothesis tests.

\subsection{Multiple Hypothesis Testing Methods}

In large-scale genomics, thousands of hypothesis tests are run in parallel, necessitating corrections to control the FDR or family-wise error rate. Here, three FDR-based methods are compared:

\begin{enumerate}
    \item \textbf{Benjamini--Hochberg (BH)}: Controls FDR under independence or certain positive correlation structures. Commonly used in many pipelines.
    \item \textbf{Benjamini--Yekutieli (BY)}: More conservative, valid under arbitrary dependency. Yields fewer declared DE genes but ensures very strict FDR control.
    \item \textbf{Storey's Q-value}: Estimates $\pi_0$, the proportion of true nulls, which can increase power if accurately estimated. Also addresses certain dependency structures.
\end{enumerate}

All methods are applied at a significance level $\alpha = 0.05$.

\section{Interpreting the Code Output and Confusion Matrix}

After applying these tests to the simulated dataset, the following performance metrics and confusion matrix results were generated:

\begin{center}
\begin{tabular}{lcccccccc}
\toprule
\textbf{Method} & \textbf{TypeI} & \textbf{FDR} & \textbf{Power} & \textbf{TP} & \textbf{FP} & \textbf{TN} & \textbf{FN}\\
\midrule
BH       & 0.0134 & 0.0330 & 0.9180 & 2754 & 94  & 6906 & 246 \\
BY       & 0.0004 & 0.0013 & 0.7507 & 2252 & 3   & 6997 & 748 \\
StoreyQ  & 0.0151 & 0.0369 & 0.9213 & 2764 & 106 & 6894 & 236 \\
\bottomrule
\end{tabular}
\end{center}

\subsection{Definitions}
\begin{itemize}
    \item \textbf{TP} (true positives): Genes truly DE and declared significant.
    \item \textbf{FP} (false positives): Genes not DE but declared significant.
    \item \textbf{TN} (true negatives): Genes not DE and not declared significant.
    \item \textbf{FN} (false negatives): Genes truly DE but missed (not declared significant).
\end{itemize}

\subsection{Type I Error, FDR, and Power}
\begin{itemize}
    \item \textbf{Type I Error} = $\frac{\mathrm{FP}}{\mathrm{FP}+\mathrm{TN}}$. BH = 1.34\%, BY = 0.04\%, StoreyQ = 1.51\%.
    \item \textbf{FDR} = $\frac{\mathrm{FP}}{\mathrm{TP} + \mathrm{FP}}$. BH = 3.30\%, BY = 0.13\%, StoreyQ = 3.69\%.
    \item \textbf{Power} = $\frac{\mathrm{TP}}{\mathrm{TP} + \mathrm{FN}}$. BH $\approx$ 91.8\%, BY $\approx$ 75.1\%, StoreyQ $\approx$ 92.1\%.
\end{itemize}

The Benjamini--Yekutieli (BY) method is the most conservative, drastically reducing false positives at the cost of lower power. BH and StoreyQ methods are less conservative (higher FP) yet maintain higher power while still controlling FDR below 5\%.

Since the simulation was designed with exactly 30\% of genes being DE, the results confirm that:

The \textbf{Benjamini-Hochberg (BH): }  Shows a Type I error of \(0.0134\), an FDR of \(0.0330\), and a statistical power of \(0.9180\). It identifies \(2754\) true positives, with \(94\) false positives and \(246\) false negatives, balancing statistical rigor with sensitivity.

The \textbf{Benjamini-Yekutieli (BY): } This is more conservative, with a Type I error of \(0.0004\), FDR of \(0.0013\), and lower power (\(0.7507\)). It detects \(2252\) true positives, with only \(3\) false positives but \(748\) false negatives, making it less suitable for exploratory analyses where capturing true positives is crucial.

\textbf{Storey’s Q-value (StoreyQ): } Strikes a balance between the two, with a Type I error of \(0.0151\), FDR of \(0.0369\), and the highest power (\(0.9213\)) among the methods. It identifies \(2764\) true positives, with \(106\) false positives and \(236\) false negatives, making it ideal for scenarios where maximizing discoveries is more important than controlling false positives.

In summary:
- \textbf{BH} is best for balanced performance between false discovery control and power.\cite{krawczyk2018fdr}
- \textbf{BY} is suited for avoiding false positives but sacrifices true positive identification.
- \textbf{StoreyQ} is optimal for maximizing significant discoveries in exploratory genomics research.

\subsection*{Conclusion:}
Throughout this research journey, I gained valuable insights into the complex world of multiple hypothesis testing, especially in the context of high-dimensional genomics data. One of the most significant takeaways was understanding the intricacies of methods like the Benjamini-Hochberg (BH) procedure, Benjamini-Yekutieli (BY) approach, and Storey’s method. These techniques are crucial for controlling false discovery rates, particularly when dealing with vast datasets where the risk of Type I errors is high.

In addition to the theoretical learning, I also honed my practical skills by learning to code in R and performing multiple hypothesis tests, which allowed me to simulate and analyze data efficiently. The hands-on experience with R not only improved my programming capabilities but also gave me a deeper understanding of how to handle complex data, visualize results, and make informed decisions about thresholding methods.

Simulating data and interpreting test results became an integral part of this journey, as it helped solidify the concepts and provided a clearer perspective on how these statistical methods can be applied in real-world genomics research. Ultimately, this experience enhanced my ability to tackle high-dimensional data challenges and laid the foundation for future exploration into more advanced methods of hypothesis testing.

This journey not only improved my technical skills but also sparked a passion for pursuing deeper research in genomics, and I look forward to building on these foundations in future studies.

\section{Extended Discussion on Multiple Testing and Statistical Implications}
\label{sec:extended_multitest}

To further elaborate on the issue of multiple testing in high-throughput analyses, we delve into theoretical and practical nuances that shape how FDR control is exercised. While the concept is deeply rooted in controlling the proportion of false positives among all declared significant findings, real-world omics data add layers of complexity.

\subsection{Correlation and Dependence Among Tests}
Genes are often co-expressed in pathways, meaning that their expression levels can be correlated, especially within a functional module. Traditional procedures like Benjamini-Hochberg (BH) assume independence or mild dependencies that do not drastically inflate false positives. However, pronounced correlation structures may require more conservative adjustments, such as Benjamini-Yekutieli (BY) . Alternatively, novel methods attempt to cluster genes by expression patterns and apply hierarchical corrections, potentially striking a balance between too-lenient assumptions and overly conservative procedures.

\subsection{Conservative vs. Adaptive Approaches}
The BY procedure, for example, replaces the denominator $G$ (the total number of tested genes) with $G \cdot \sum_{i=1}^G \frac{1}{i}$ (the harmonic sum), ensuring valid FDR control even under dependencies. This can drastically reduce power if the dataset does not genuinely have strong correlations. On the other hand, Storey’s q-value approach tries to adapt to the data by estimating the proportion of truly null hypotheses ($\pi_0$). When $\pi_0$ is significantly less than 1, this method can afford to be more permissive while still controlling the overall FDR. Such adaptivity can recapture lost power in large-scale genomics assays.

Ultimately, a nuanced approach to multiple testing acknowledges the synergy between statistical principles and domain-specific knowledge. Researchers remain encouraged to incorporate pilot data, replicate findings in independent cohorts, or rely on complementary assays to ensure that reported genes truly represent meaningful biological differences.

\section*{Limitations}

\begin{itemize}
    \item \textbf{Simulated Data Assumptions:} The analysis primarily relied on simulated data, which may not fully capture the complexities and noise present in real-world RNA-seq datasets.
    \item \textbf{Method Dependency:} The effectiveness of BH, BY, and Storey’s methods can vary based on the specific characteristics of the dataset, such as the correlation structure between genes and the number of tests performed.
    \item \textbf{Batch Effects:} Potential batch effects were not fully addressed in the analysis, which may lead to biased results when applying the methods to real-world data.
    \item \textbf{Gene-Gene Correlations:} Correlations among genes were assumed to be simple, but in reality, gene interactions can be highly complex,\cite{reiner2003multiple} affecting the accuracy of multiple testing corrections.
    \item \textbf{Generalization:} The methods discussed may not generalize well to all types of high-dimensional data, particularly in cases of sparse or unbalanced datasets.
    \item \textbf{Computational Complexity:} The computational requirements for performing these multiple testing corrections on very large datasets can be high, limiting the scalability of the analysis.
\end{itemize}
Such limitations  highlight the need for further research and exploration to address these challenges and enhance the robustness of differential expression analysis in genomics.

\printbibliography

\end{document}